\documentclass[prb,superscriptaddress,floatfix,a4paper,aps,twocolumn,nofootinbib]{revtex4-2}

\usepackage{graphicx}
\usepackage{amsmath}
\usepackage{amssymb}
\usepackage[usenames, dvipsnames]{xcolor}
\usepackage[colorlinks=true,citecolor=blue,linkcolor=blue,urlcolor=blue]{hyperref}

\usepackage{tabularx}
\usepackage{color}
\usepackage{xspace}
\usepackage{hyperref}

\begin{document}
		
	\title {Electronic band structure from quasiparticle interference and Landau quantization in WTe$_2$}%
	
	\author{Raquel Sánchez-Barquilla}
	\affiliation{Laboratorio de Bajas Temperaturas y Altos Campos Magnéticos, Departamento de Física de la Materia Condensada, Instituto Nicolás Cabrera and Condensed matter Physics Center (IFIMAC), Unidad Asociada UAM-CSIC, Universidad Autónoma de Madrid, E-18049, Madrid, Spain}
		
	\author{Francisco Martín Vega}
	\affiliation{Laboratorio de Bajas Temperaturas y Altos Campos Magnéticos, Departamento de Física de la Materia Condensada, Instituto Nicolás Cabrera and Condensed matter Physics Center (IFIMAC), Unidad Asociada UAM-CSIC, Universidad Autónoma de Madrid, E-18049, Madrid, Spain}
	
	\author{Alberto M. Ruiz}
	\affiliation{Instituto de Ciencia Molecular (ICMol), Universidad de Valencia, Catedrático José Beltrán 2, 46980 Paterna, Spain}

	\author{Na Hyun~Jo$^*$}
	\affiliation{Ames Laboratory and Department of Physics \& Astronomy, Iowa State University, Ames, IA 50011}
\altaffiliation{Present address: Department of Physics, University of Michigan, Ann Arbor, MI 48109, USA}

		\author{Edwin Herrera}
	\affiliation{Laboratorio de Bajas Temperaturas y Altos Campos Magnéticos, Departamento de Física de la Materia Condensada, Instituto Nicolás Cabrera and Condensed matter Physics Center (IFIMAC), Unidad Asociada UAM-CSIC, Universidad Autónoma de Madrid, E-18049, Madrid, Spain}
	
		\author{José J. Baldoví}
	\affiliation{Instituto de Ciencia Molecular (ICMol), Universidad de Valencia, Catedrático José Beltrán 2, 46980 Paterna, Spain}
	
	\author{Masayuki~Ochi}
    \affiliation{Forefront Research Center, Osaka University, Toyonaka, Osaka 560-0043, Japan} 
	\affiliation{Department of Physics, Osaka University, Toyonaka, Osaka 560-0043, Japan}
	
	\author{Ryotaro~Arita}
	\affiliation{RIKEN Center for Emergent Matter Science, Wako, Saitama 351-0198, Japan}
	
	\author{Sergey L. Bud'ko}
	\affiliation{Ames Laboratory and Department of Physics \& Astronomy, Iowa State University, Ames, IA 50011}
	
	\author{Paul C. Canfield}
	\affiliation{Ames Laboratory and Department of Physics \& Astronomy, Iowa State University, Ames, IA 50011}
		
	\author{Isabel Guillamón}
	\affiliation{Laboratorio de Bajas Temperaturas y Altos Campos Magnéticos, Departamento de Física de la Materia Condensada, Instituto Nicolás Cabrera and Condensed matter Physics Center (IFIMAC), Unidad Asociada UAM-CSIC, Universidad Autónoma de Madrid, E-18049, Madrid, Spain}
	
	\author{Hermann Suderow}
	\affiliation{Laboratorio de Bajas Temperaturas y Altos Campos Magnéticos, Departamento de Física de la Materia Condensada, Instituto Nicolás Cabrera and Condensed matter Physics Center (IFIMAC), Unidad Asociada UAM-CSIC, Universidad Autónoma de Madrid, E-18049, Madrid, Spain}	

	\date{\today}
	
	\begin{abstract}
WTe$_2$ stands out as a semimetal presenting Fermi level quantum oscillations in most measured quantities under magnetic fields. However, the electronic band structure above and below the Fermi level has not been explored completely. Here we study the electronic band structure of WTe$_2$ by quasiparticle interference with Scanning Tunneling Microscopy (STM) and observe, with the support of Density Functional Theory (DFT), the electron and hole bands around the Fermi level. We also report on the observation of Landau quantization in atomically resolved measurements and discuss the possible connection with band structure calculations.
	\end{abstract}
	
	\maketitle
	
	\section{Introduction}

The bulk boundary correspondence might lead to topologically protected surface states in materials presenting band inversion in the band structure\,\cite{RevModPhys.82.3045,RevModPhys.83.1057,Wieder2022,Nagaosa2020,annurev:/content/journals/10.1146/annurev-conmatphys-031016-025458,RevModPhys.90.015001,RevModPhys.93.025002,PhysRevB.85.035135,PtSn4wu2016,Inoue2016}. Experiments and band structure calculations show that the semimetal WTe$_2$ presents tilted conical crossings created by band inversion. These lead to Weyl fermions, which are connected by Fermi arcs at the surface. The Fermi arcs are expected to arise within a narrow energy interval close to the Fermi level\,\cite{Weyl1929,ali2014,Das_2019,xu2015,soluyanov2015,wu2016,wu2015,wang2016,PhysRevB.94.161401,bruno2016,zhang2017,yuan2018,Pan2015,Kang2015,PhysRevLett.127.257401,doi:10.1126/science.aar4426,Maximenko2022,Hwang2020}. Angle-resolved photoemission (ARPES) experiments revealed surface arc-like bands spanning a broad range in energy, and were initially interpreted as trivial surface states\,\cite{wu2016,yuan2018,Kawahara_2017,wang2016,bruno2016,wu2015}. Other experiments demonstrated by contrast non-trivial topological characteristics, such as the observation of spinful hinge states\,\cite{Kononov2020,Choi2020,Lee2023}, a hallmark of higher order topological insulators. WTe$_2$ was proposed to be near a higher order topological insulator phase due to double band inversion (similar to bismuth\,\cite{Schindler2018,doi:10.1126/sciadv.aat0346,PhysRevLett.123.186401}) and the large surface states with linear dispersion previously found by ARPES to be remnant surface states arising from the proximity to the higher order topological insulator\,\cite{PhysRevLett.123.186401}.

On the other hand, WTe$_2$ stands out among the layered dichalcogenides because it is easy to observe quantum oscillations due to Fermi level Landau quantization in nearly all macroscopic properties measured under magnetic fields\,\cite{ali2014,PhysRevLett.114.156601,zhu2015}. Scanning Tunneling Microscopy (STM), which measures the local electronic density of states at the surface as a function of the energy by varying the bias voltage between tip and sample, seems an ideal tool to study Landau quantization above and below the Fermi level. The establishment of well-defined Landau levels is more favorable when there is no dispersion along the axis parallel to the magnetic field\,\cite{annurev-conmatphys-040721-021331}. Landau quantization observed by STM has thus been used to characterize two-dimensional systems as graphene and two-dimensional electron gases, and surface states of bismuth and several semiconductors, as Bi$_2$Se$_3$\,\cite{doi:10.1126/science.1171810,doi:10.1126/science.aag1715,PhysRevB.82.081305,doi:10.1126/science.1239451,PhysRevB.107.115426,PhysRevLett.118.016803,Tsui2024,Yin2016,PhysRevB.92.165420}. Until now, and to the best of our knowledge, Landau quantization with STM has not been reported in WTe$_2$\,\cite{zhang2017,yuan2018,das2016, pletikosic2014,Maximenko2022,Hwang2020}.

In this study, we present a detailed investigation of the electronic local density of states of WTe$_2$ at zero magnetic field, enabling us to determine both bulk and surface band structures in the vicinity of the Fermi energy. We also demonstrate the presence of Landau quantization through measurements conducted under high magnetic fields, reaching up to 14 T.

\begin{figure}
	\includegraphics[width=0.86\columnwidth]{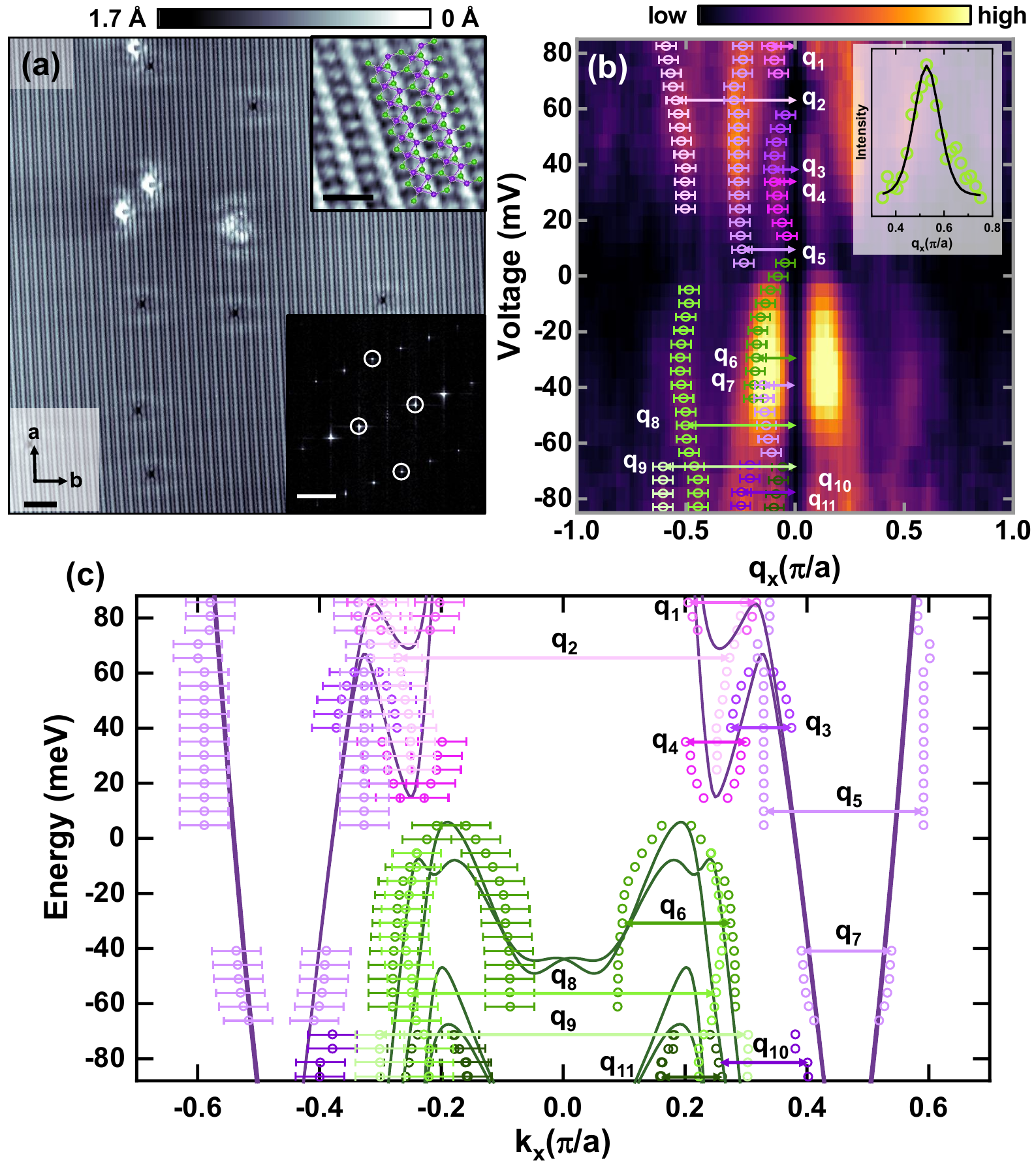}
	\caption{(a) Scanning Tunneling Microscope (STM) topography of the WTe$_2$ surface, revealing parallel Te rows. The image encompasses various types of defects, around which wave-like patterns are observed, primarily attributed to the quasi-one-dimensional band dispersion along the Te rows. The upper right inset presents a magnified view of a defect-free region, with Te atoms depicted in green and the underlying W atoms in violet. The lower right inset displays the Fourier transform of the main panel image, where white circles highlight the Bragg peaks corresponding to the atomic lattice. A more detailed description of the crystal structure of WTe$_2$ and the observed defects can be found in Appendix, Section A. The black horizontal scale bar represents 2 nm. The grayscale bar at the top represents the height changes in the main panel. In the upper right inset height changes are of about 0.6\,\AA\, and the horizontal scale bar is 0.6 nm long. The white scale bar in the lower right inset is 2 nm$^{-1}$ long. (b) Fourier transform of the tunneling conductance maps, displayed as a color scale (see top bar), plotted as a function of bias voltage and the wave vector along the Te chains ($q_x$, parallel to the crystal's a-axis). Colored points and arrows, labeled $q_i$ ($i=1...11$), indicate identified scattering wave vectors. Data along other in-plane directions, including a feature at large $q$ which remains unmarked here, are presented and discussed in Appendix, Section B. In the upper right inset we show as green circles the intensity of the Fourier transform at about $-50$ mV related to the feature leading to the $q_8$ vector. Black line is a Gaussian fit to the data. (c) Density Functional Theory (DFT) band structure calculations of bulk WTe$_2$ are shown as colored lines, with W-derived bands in violet and Te-derived bands in green. Error bars are obtained from fitting the quasiparticle scattering intensity as a function of $q$ with a Gaussian. We show one of these Gaussian fits as a black line in the inset of (b). Error bars in (b,c) are given by the half-width of the Gaussian. The colored points and arrows, $q_i$, correspond to the experimental scattering patterns identified in (b).}
	\label{FigTopo}
\end{figure}

	\section{Experiments and calculations}
	
We performed STM measurements using a home-made set-up described in Ref.\,\cite{suderow2011} and the software described in Ref.\,\cite{fran2021}. Images and spectroscopy data were processed using the software described in Refs.\,\cite{fran2021, horcas2007}. The samples were grown from a Te-rich binary melt, as described in Refs.\, \cite{canfield1992, wu2015}. The crystals (of the 1T$_d$ polytype) were plate-like with typical dimensions of 2 mm $\times$ 0.5 mm $\times$ 0.05 mm, and the crystallographic $c$-axis was perpendicular to the crystal surface. We exfoliated the sample in-situ in cryogenic conditions by gluing a post and pushing it at low temperatures using the in-situ slider described in Ref.\,\cite{suderow2011}. This method provided large, clean and flat surfaces, showing atomic rows and atomic size defects, similar to those reported before\,\cite{zhang2017, yuan2018, das2016, pletikosic2014,Maximenko2022}. Experiments were made in a helium bath cryostat at 4.2\,K and in superconducting coils from Oxford Instruments and Cryogenics, delivering fields of 14 T. We succesfully cleaved eight crystals, in which we changed the field of view more than a hundred times. We used a tunneling current of 4 nA and most often bias voltages of 100 mV. We took tunneling conductance curves as usual, by cutting the feedback loop and sweeping the bias voltage. We derived numerically and smoothed the resulting current vs voltage traces, using similar methods as in previous work\,\cite{suderow2011,fran2021,PhysRevB.77.134505}. The resolution in bias voltage was of about 2 mV, similar to the one obtained previously\,\cite{zhang2017,yuan2018,das2016, pletikosic2014,Maximenko2022,Hwang2020}. When plotting the Fourier transform, we remove the large contribution at low wavevectors (below about 0.1 $\pi/a$, we use a Gaussian filter centered at zero wavevector) to allow for the establishment of a color scale that highlights the scattering features. This eliminates the contributions from large features, as impurities. We furthermore symmetrize the Fourier transform using the mirror planes of the electronic band structure. Using the in-situ sample positioning system\,\cite{suderow2011}, we studied fields of view with defects and impurities, where we took the quasiparticle interference (QPI) data, and others completely free of impurities, where we observed Landau quantization.

First-principles band structure calculations were performed based on DFT with the generalized gradient approximation with the Perdew-Burke-Ernzerhof parametrization~\cite{Perdew1996} and the full-potential (linearized) augmented plane-wave plus local orbitals method including the spin-orbit coupling as implemented in the \textsc{WIEN2k} code~\cite{Blaha2018,Blaha2020}, using the crystalline structure of Ref.~\cite{Mar1992}. The maximum modulus for the reciprocal lattice vectors $K_{\mathrm{max}}$ was chosen so that $rK_{\mathrm{max}}=8$, where the muffin-tin radius $r$ was 2.5 Bohr for both W and Te atoms. We also calculated the surface states (see Fig.\,\ref{FigQuantumnumber}(e) and Fig.\,\ref{QPIComparison} in the Appendix B), for which we used the semi-infinite-slab tight-binding model described in Ref.~\cite{Sancho1985}.

	\section{Results}
	
	\subsection{Quasiparticle interference}

The atomic structure observed at the surface is characterized by topmost layers where Te atoms form the primary atomic lattice. Our STM images resolve these Te atomic positions and, through their arrangement, reveal the underlying W atoms located in between (see the top right inset of Fig.\,\ref{FigTopo}(a)). The prepared surfaces exhibit a remarkably low defect density, predominantly consisting of W-Te substitutions and vacancies. As detailed in Appendix Section A, the defect concentration is approximately one defect per 700 atoms, translating to around $10^{12}$ defects per cm$^{-2}$. This exceptionally low defect density is consistent with the extremely high residual resistance ratio reported for these crystals and the clear observation of quantum oscillations in macroscopic experiments\,\cite{wu2015,wu2016,jo2019}.

Defects on the WTe$_2$ surface act as scattering centers, generating the oscillatory patterns in the local density of states $N(E)$ that are revealed through QPI imaging. To effectively study these scattering patterns, we specifically targeted fields of view exhibiting a sufficient density of impurities (see Appendix Section B, Fig.\,\ref{ConducMaps}). The results of our QPI measurements are presented in Fig.\,\ref{FigTopo}(b), where we show the Fourier transform of the spatial variations in the tunneling conductance along $q_x$ and as a function of the bias voltage. We then compare this experimental scattering pattern with the electronic band structure obtained from Density Functional Theory (DFT) calculations in Fig.\,\ref{FigTopo}(c), allowing us to identify the wave vectors ($\mathbf{q}$) associated with electronic scattering processes. The scattering vectors along $q_x$ connect portions of the band structure along the direction of $k_x$. In Fig.\,\ref{FigTopo}(b,c) we focus on wave vectors at $k_y=0$. There is intraband scattering between states at $\pm k_x$,  resulting in a scattering vector $q_x=2k_x$ (see for example the light pink arrow $\mathbf{q}_2$ in Fig.\,\ref{FigTopo}(b,c)). Similarly, intraband scattering across a single Fermi pocket with a size $\delta k_x$ yields a scattering vector $q=\delta k_x$ (as shown by the dark green arrows $\mathbf{q}_6$ in the lower middle of Fig.\,\ref{FigTopo}(c)). Further details on the scattering vectors are provided in Appendix, Section B. Notably, we observe scattering patterns that are elongated along the a-direction (labeled as $x$ in Fig.\,\ref{FigTopo}), which corresponds to the direction of the Te chains (as visualized in Fig.\,\ref{FigTopo}(a); see Appendix, Fig.\,\ref{QPIComparison}, for QPI results along other in-plane directions).

\begin{figure}
	\includegraphics[width=1\columnwidth]{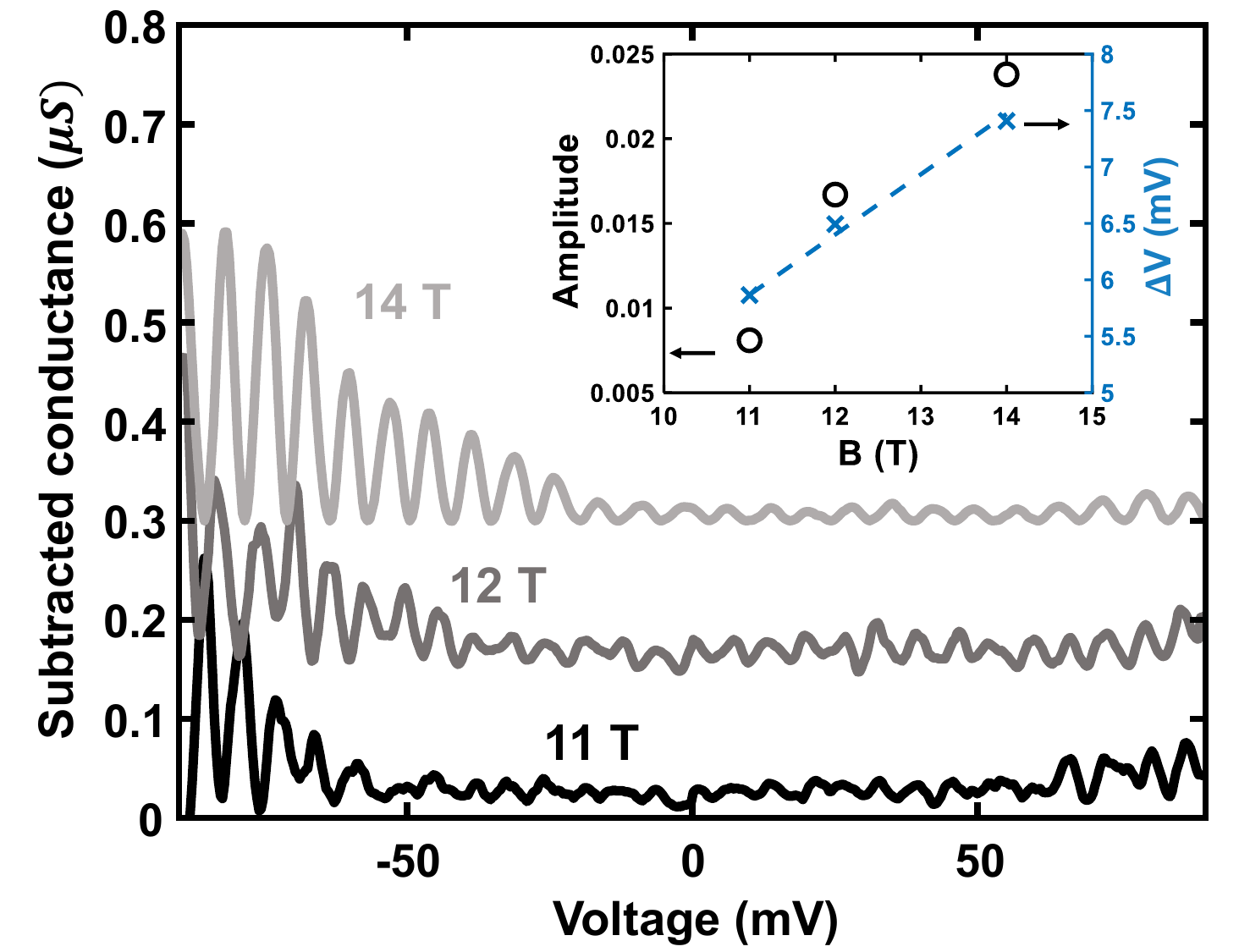}
	\caption{Tunneling conductance as a function of bias voltage measured at 11\,T, 12\,T, and 14\,T. To enhance visibility, a background signal has been subtracted from each curve, and the resulting data have been vertically shifted. The background is similar to the tunneling conductance curves shown in Fig.\,\ref{BandStructure}(b) and has a V-shape often found in semimetals (see Fig.\,\ref{Background} for the subtraction). The amplitude of the observed oscillations under magnetic fields is up to 40\% of the tunneling conductance. The inset displays the amplitude of the oscillatory component of the conductance as circles, while the energy separation between adjacent oscillation peaks is shown as crosses. The dashed line in the inset represents the usual magnetic field dependence of the Landau level energy separation for an effective mass of  $m^* = 0.22 m_e$.}
	\label{FigLandau}
\end{figure}

In Fig.\,\ref{FigTopo}(c), we compare our DFT calculations (lines) with the experimental scattering vectors obtained from QPI measurements (points). At large positive bias voltages, we observe both a small and a large scattering vector (magenta and light magenta arrows, $\mathbf{q}_1$ and $\mathbf{q}_2$ in Fig.\,\ref{FigTopo}(c)). The smaller vector $\mathbf{q}_1$ rises from intraband scattering within a small W-derived electron pocket, while the larger vector $\mathbf{q}_2$ corresponds to intraband scattering between the outer edges of another, larger electron pocket (as visualized in Fig.\,\ref{FigTopo}(b)). Similarly, at negative bias voltages, focusing on the Te-derived bands (shown in green in Fig.\,\ref{FigTopo}(c)), we identify a scattering vector (green arrow $\mathbf{q}_{6}$) originating from intraband scattering between two distinct hole bands. We also detect scattering vectors associated with interband transitions between W-derived electron bands and Te-derived hole bands (bright violet $\mathbf{q}_{10}$ in Fig.\,\ref{FigTopo}(b,c)). Overall, the experimental QPI data show a good agreement with the calculated band structure, revealing a complex band structure composed of predominantly W-derived electron pockets and Te-derived hole pockets with intricate shapes.

The band structure we determined below the Fermi level aligns well with previous ARPES studies of occupied states\,\cite{pletikosic2014,wu2015,wu2016}. Importantly, our QPI measurements further reveal the mixed W-derived electron and Te-derived hole character of specific bands, as evidenced by the scattering vectors $\mathbf{q}_{6}-\mathbf{q}_{11}$. Furthermore, our study extends the understanding of the electronic structure by determining the band dispersion above the Fermi level. We observe that the W-derived electron-like bands exhibit hole-like character within a small energy range (indicated by the scattering wavevector $\mathbf{q}_3$ in Fig.\,\ref{FigTopo}(c)) above the Fermi level. A more detailed comparison between our theoretical calculations and experiments, including results along directions away from the $a$-axis is presented in the Appendix, Section B and Figs.\,\ref{ConducMaps},\ref{QPIComparison}. As we show below and in the Appendix Section B (orange lines in Fig.\,\ref{QPIComparison}), there are additional scattering features that can be related to surface quasi-two-dimensional bands which are also compatible with previous ARPES studies.

Our QPI results generally compare well with previous QPI studies\,\cite{zhang2017,yuan2018,Hwang2020}, which show a similar set of scattering vectors around the Fermi level for $q_y=0$ as in Fig.\,\ref{FigTopo}(b,c) (albeit sometimes shifted in bias voltage) and similar patterns for finite $q_y$, including several arc-like features discussed in the Appendix, Section B and Figs.\,\ref{ConducMaps},\ref{QPIComparison}. Here, we obtain a more detailed comparison with calculations, Fig.\,\ref{FigTopo}(c), and identify the set of electron and hole bands across the Fermi level.

\begin{figure*}
	\includegraphics[width=1\textwidth]{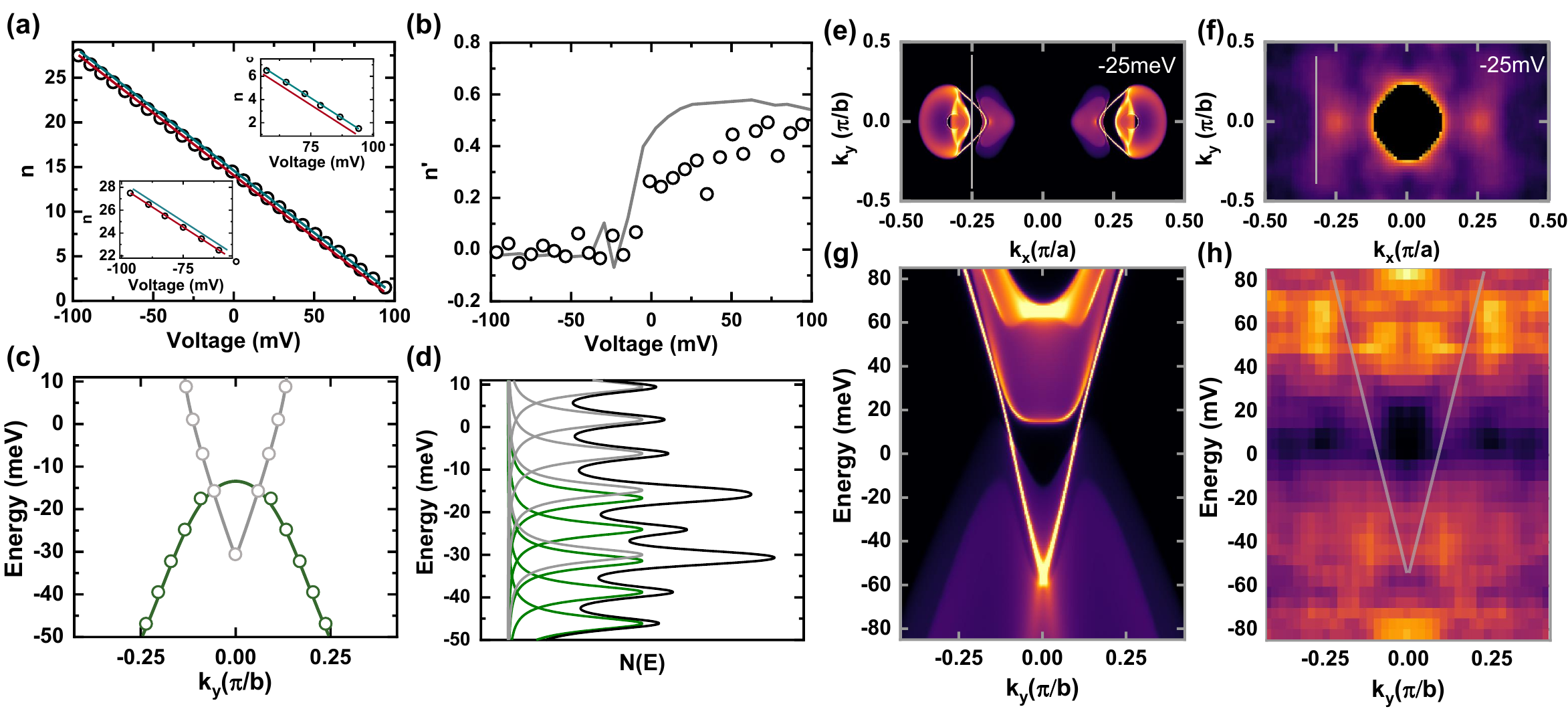}
	\caption{(a) The sequence of peaks in the tunneling conductance vs bias voltage is plotted as circles for a magnetic field of 14\,T. Each peak is labelled by a number $n$. The blue line represents a linear fit to the peak sequence for positive bias voltages, while the red line shows the linear fit for negative bias voltages. The upper right and lower left insets provide magnified views of the data at high and low bias voltages. (b) The deviation ($n'$) of the observed peak index (from panel a) from the linear fit for negative bias voltages (red line in a) is shown as circles. The grey line illustrates the peak count predicted by considering the combined linear and parabolic band structure detailed in (c, d), and obtained as described in the text. (c) Minimal model providing a shift in the Landau level sequence of two-dimensional bands. Schematic representation of a linearly dispersing band (grey line) intersecting with a closely lying parabolic band (green line). The circles schematically depict the resulting Landau levels formed in a magnetic field. Further details on this model are provided in Appendix Section C. (d) Calculated density of states $N(E)$ as a function of energy, incorporating energy broadening for both the linear (light grey) and parabolic (green) bands shown in (c). The total $N(E)$ is represented by the dark grey line. The energy shift observed at the overlap between the linear and parabolic bands (between -15 meV and -30 meV) leads to a corresponding shift in the Landau level count, as demonstrated in (b).  (e) Constant energy contours at the surface obtained from a semi-infinite slab calculation, providing the surface band structure at an energy of -25 meV, as a function of in-plane wave vectors $k_x$ and $k_y$. The intensity of $N(E)$ is indicated by the color scale, ranging from black (low) to yellow (high). (f) Experimentally measured scattering pattern at a bias voltage of -25 mV.  (g) Calculated surface band structure along the grey line indicated in (e). There is a V-shape structure exhibiting cuasi linear dispersion, surrounded in particular by a hole surface band with parabolic dispersion. This motivated the minimal model of (c,d). (h) QPI pattern at the same reciprocal space region as (g). The conversion of the experimental scattering pattern to wave vectors $k_x$ and $k_y$ assumes intraband scattering around $\pm k_y=0$ and is obtained simply by dividing the experimental scattering pattern by two. Grey lines are at the same position as the V-shaped structure in (g).}
		\label{FigQuantumnumber}
\end{figure*}

	\subsection{Landau quantization}

While defects can generally hinder the observation of Landau quantization \cite{peng2010}, a crucial length scale for Landau level formation is the magnetic length, $\ell_B = \sqrt{\hbar/eB}$, which also dictates the radius of the $n-$th cyclotron orbit $r_n = \sqrt{2n+1}\ell_B$. To resolve discrete Landau levels in the density of states $N(E)$, the average distance between scattering defects should ideally be comparable to or larger than the magnetic length, $\ell_B$. At our maximum field of 14 T, the radius of the lowest Landau level orbit ($r_0 \approx \ell_B \approx$ 7 nm) is significantly smaller than the estimated mean free path $\ell$ (of about $\ell\approx 170$ nm, see Appendix B). This suggests that Landau quantization occurs under a magnetic field in our samples. However, the tunneling conductance obtained by spatially averaging measurements over large, nominally impurity-free regions (see Appendix, Section D, Fig.\,\ref{BandStructure}(b)) shows no significant dependence on the applied magnetic field and, importantly, no signatures of Landau quantization.

Landau quantization modifies the constant energy countours of the band structure into degenerate Landau circles (for two dimensional surface bands) or tubes (for three dimensional bulk bands) along the direction of the magnetic field. The in-plane band structure is quantized, whereas the out-of-plane band structure remains dispersive. In our STM experiments we measure the tunneling conductance as a function of the bias voltage at a fixed magnetic field. The tunneling conductance is proportional to the density of states as a function of the energy, $dI/dV(eV)\propto N(E)$. In three dimensions, each Landau tube with index $n$ is degenerate and contains the states located between $k_n$ and $k_{n+1}$ at zero magnetic field. The density of states at the Fermi level presents a peak each time a Landau tube crosses an extremal area of the Fermi surface\,\cite{annurev-conmatphys-040721-021331}. In two dimensions, by contrast, any constant energy contour of the band structure consists of a two-dimensional contour line, generally with circular shape. Under magnetic fields, the two-dimensional density of states vanishes except at energies which coincide with a Landau level, forming Landau circles only at the Landau level energy\,\cite{peng2010,doi:10.1126/science.1171810,doi:10.1126/science.aag1715,PhysRevB.82.081305,doi:10.1126/science.1239451,PhysRevB.107.115426,PhysRevLett.118.016803,Tsui2024,Yin2016,PhysRevB.92.165420,annurev-conmatphys-040721-021331}. Contrasting the situation in three dimensions, where $N(E)$ is modulated by the passage of tubes through extremal cross sections of the Fermi surface, $N(E)$ in two dimensions has peaks each time $E$ coincides with a Landau level. Such peaks are generally stronger than the peaks in $N(E)$ in three dimensional bands\,\cite{peng2010}.

In contrast to the spatially averaged conductance, when we examine tunneling conductance spectra taken at a specific, atomically defined location under magnetic fields, we observe pronounced oscillations as a function of the bias voltage (Fig.\,\ref{FigLandau}). The amplitude of these oscillations grows with increasing magnetic field, and their period in bias voltage ($\Delta V =\Delta E /e$) remains constant for a given field but increases with the magnetic field itself (inset of Fig.\,\ref{FigLandau}).

To gain deeper insight into the magnetic field behavior, we analyzed the sequence of peaks in the tunneling conductance as a function of bias voltage. We label the peaks in the conductance by $n$, starting from the lowest bias voltage where we take the tunneling conductance curve. We observe a linear dependence of $n$ with bias voltage, as expected for Landau quantization. However, our data also reveal a shift of approximately half an integer $\frac{1}{2}$ in the peak count (Fig.\,\ref{FigQuantumnumber}(a), best observed when subtracting a linear voltage dependence to obtain $n'$ shown in Fig.\,\ref{FigQuantumnumber}(b)).

To discuss this shift, we build a minimal model which can produce a shift in the Landau level sequence.

First, we consider a single parabolic surface band. The Landau level contribution to the density of states (in the absence of broadening) can be expressed as $N(E>E_{0,parab})= \sum_n \delta((E-E_{0,parab})-(n+\frac{1}{2})\hbar\omega_c)$ for the Landau level contribution, where $E_{0,parab}$ is the band top and $n$ the integer Landau level index. The tunneling conductance $dI/dV(eV)$ is then proportional to $\sum_n \int G(E,eV)  \delta((E-E_{0,parab})-(n+\frac{1}{2})\hbar\omega_c) dE$, where $G(E,eV)$ represents a broadening function due to finite temperature, experimental broadening and impurity scattering. $dI/dV$ then shows a sequence of broadened peaks separated by $\Delta V=e\Delta E =e\hbar\omega_c$. For a fixed magnetic field, the separation between Landau levels $\Delta E$ does not change with bias voltage. Therefore, if one indexes the Landau level peaks, the index $n$ increases linearly with $E$. There is no shift the level count with energy.  If the function $G(E,eV)$ provides an energy broadening which is smaller but of the order of the Landau level separation, $e\hbar\omega_c$ the index increases linearly with $E$ too. Importantly, this holds even when considering multiple contributions from different electron and hole bands with different maxima and minima.

If instead we consider a surface band with a linear dispersion, we find that the density of states contribution is $N(E>E_{0,lin})=\sum_m \delta(\vert E-E_{0,lin}\vert +\sqrt{2m\hbar e B v_F^2})$, with $v_F$ being the Fermi velocity and $m$ the Landau level index. The corresponding tunneling conductance becomes $dI/dV(eV)\propto \sum_{n,m} \int G(E,eV) (\delta((E-E_{0,lin})-(n+\frac{1}{2})\hbar\omega_c) + \delta(\vert E-E_{0,lin}\vert -\sqrt{2m\hbar e B v_F^2})) dE$. While the parabolic band yields equally spaced Landau levels, the linear dispersing band introduces a strongly energy-dependent level separation close to $E_{0,lin}$\,\cite{annurev-conmatphys-040721-021331,PhysRevLett.116.236401,PhysRevB.97.144422}.

We find that the superposition of broadened Landau levels from two surface two-dimensional bands, one with a parabollic dispersion and another one with a linear dispersion (Fig.\,\ref{FigQuantumnumber}(c); further details in Appendix, Section C) results in the a sequence of Landau levels with energy which exhibits a shift as shown schematically in Fig.\,\ref{FigQuantumnumber}(d). The corresponding expected sequence of peaks in the tunneling conductance matches the observations (Fig.\,\ref{FigQuantumnumber}(b)). We note that the linear portion of the dispersion can be located on a very small interval in energy, and might arise from a non-dominant feature in the band structure.

It is now worth checking if we can eventually identify such a sequence of a linear dispersion and a parabollic band in the QPI experiments and the band structure calculations.

In Figs.\,\ref{FigQuantumnumber}(e,g) we show the surface band structure obtained from a semi-infinite slab calculation. Note that it is a cut for along the $y$ direction at a specific $k_x$ wave vector ($k_x=-0.25$ $\frac{\pi}{a}$, grey line in Fig.\,\ref{FigQuantumnumber}(e)). When tracing the energy dependence of the band structure along this cut (Fig.\,\ref{FigQuantumnumber}(g)) we observe a linearly dispersing surface band and a hole-like surface band with parabollic dispersion.  These states have been discussed in previous ARPES experiments\,\cite{wu2016,yuan2018,Kawahara_2017,wang2016,bruno2016,wu2015} and observed in QPI studies\,\cite{Hwang2020}. We show our QPI data in the same region of reciprocal space in Fig.\,\ref{FigQuantumnumber}(h). To obtain the wave vector range in reciprocal space from the QPI experiment, we assume intraband scattering and obtain the momenta $k_y$ by halving the scattering vector $q_y$. There are hints in the QPI for features which are comparable to calculations of the surface band structure. In particular, there is a broad background at positive and negative bias voltages, and a feature at a finite $k_y$ close to zero bias. Close to zero bias, the passage of the grey line roughly coincides with a feature in the scatterin pattern. This feature mixes with further surface and bulk scattering features, discussed in detail in the Appendix, Section B Fig.\,\ref{QPIComparison}.
	
An intriguing additional observation is that the the sequence of peaks in the tunneling conductance under magnetic fields is not uniform across the surface but varies with the atomic position. The background tunneling conductance (Fig.\,\ref{BandStructure}) remains roughly independent of atomic position. However, the peaks in the tunneling conductance (Fig.\,\ref{FigLandau}) have a strong spatial dependence. This spatial dependence gives rise to patterns in the tunneling conductance maps. We can find a minimal model to explain this feature by incorporating a spatially dependent shift in the Landau level indexing, as shown with detail in Appendix, Section E and Fig.\,\ref{FigConductanceMap}. The spatially dependent shift could be caused if tunneling onto different atoms is dominated by contributions from bands with different character on different positions. This provides an explanation for the absence of well-defined peaks under magnetic fields in the spatially averaged density of states. The rapidly varying spatial peak sequence is smeared out when averaging over larger areas, leading to a tunneling conductance which is smooth as a function of the bias voltage.

\section{Conclusions}

The identification of Landau-quantized quasiparticles  in our experiment suggests that there could be quasi-two-dimensional surface bands. These could play a role in the behavior observed in transport experiments on few-layer WTe$_2$ devices and heterostructures\,\cite{PhysRevB.107.245410,Fei2017,PhysRevB.96.041108,doi:10.1126/science.aan6003,Tang2017,PhysRevB.105.094512,doi:10.1021/acs.nanolett.8b03924,Shi2019,PhysRevB.105.094512,doi:10.1021/acs.nanolett.8b03924,Shi2019,Choi2020,PhysRevB.98.075430}.

In summary, using QPI, we have comprehensively mapped the electronic band structure of WTe$_2$ in the vicinity of the Fermi level, revealing both electron-like and hole-like bands above and below the Fermi level. We have also discussed the presence of Landau-quantized levels when measuring with atomic precision. The distinction between atomically resolved and averaged tunneling conductance measurements could be of help to identify Landau quantization in other materials with a complex bulk band structure.

\section{Acknowledgments}

This work was supported by the Spanish Research State Agency (PID2023-150148OB-I00, CEX2023001316-M, TED2021-130546B-I00, PDC2021-121086-I00), by the Comunidad de Madrid through project TEC-2024/TEC-380 “Mag4TIC”, by the European Research Council PNICTEYES grant agreement 679080, and by VectorFieldImaging grant agreement 101069239. We acknowledge collaborations through EU program Cost CA21144 (Superqumap). We acknowledge segainvex for design and construction of STM and cryogenic equipment. Work at Valencia received support from the European Union (ERC-2021-StG-101042680 2D-SMARTiES) and the Gen-T programme of the Generalitat Valenciana (CIDEXG/2023/1). A.M.R. thanks the Spanish MIU (FPU21/04195). Work at the Ames National Laboratory was supported by the U.S. Department of Energy, Office of Science, Basic Energy Sciences, Materials Sciences and Engineering Division. Ames National Laboratory is operated for the U.S. Department of Energy by Iowa State University under Contract No. DE-AC02-07CH11358.

\newpage

\section*{Appendix}

\subsection{Atomic structure of the surface of WTe$_2$}

\begin{figure*}
	\includegraphics[width=1\textwidth]{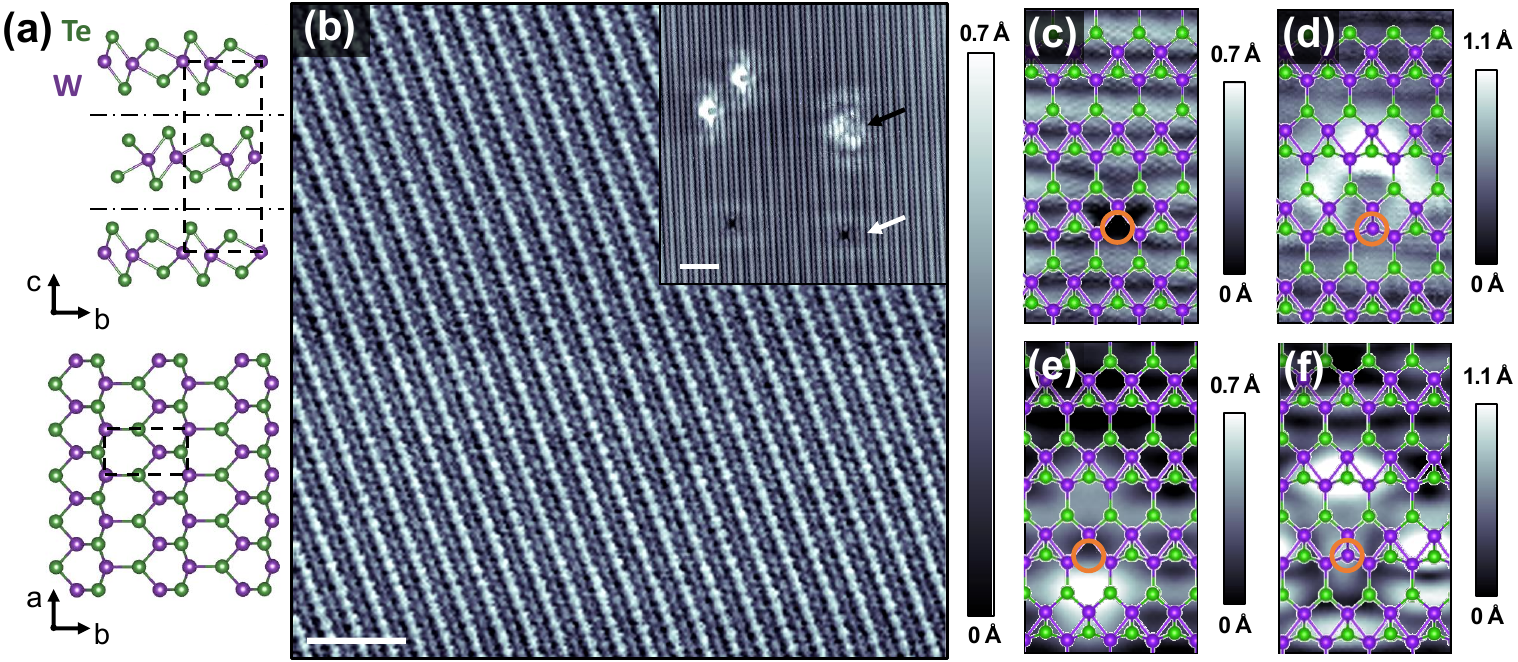}
	\caption{(a) Crystal structure of 1T$_d$ WTe$_2$. The top panel presents a side view, revealing the layered orthorhombic structure and the van der Waals gap (dashed-dotted lines) between adjacent layers. Tungsten (W) atoms are depicted as violet disks, and tellurium (Te) atoms as green disks. The bottom panel shows a top-down view, illustrating the in-plane arrangement of W and Te atoms and the corresponding in-plane unit cell (dashed lines). The unit cell dimension along the stacking c-axis is indicated by dashed lines in the top panel. (b) Large-area STM topography of WTe$_2$. This image shows a defect-free region of the WTe$_2$ surface within the field of view. The inset provides a zoomed-in view of a different area (shown in Fig.\,\ref{FigTopo}(a)), highlighting a tungsten interstitial (black arrow) and a tellurium vacancy (white arrow). The white horizontal scale bar in the main panel and in the inset represents 2 nm. The color scale on the right indicates the height variations, ranging from white (highest) to black (lowest). (c,d) Topographic STM images around defects. We superpose the atomic pattern of the surface. We show by an orange circle the position of a missing Te atom (c) and a W interstitial (d). (e) Calculated local density of states on a Te vacancy and in (f) for a W interstitial. Note that the maps (c-f) are rotated with respect to (b) so as to place the Te chains parallel to the x-axis.}
	\label{TopoTeo}
\end{figure*}

WTe$_2$ is a transition metal dichalcogenide characterized by a layered crystalline structure. Typically, these materials feature a layer where a transition metal is coordinated between two chalcogen atoms. In WTe$_2$, each Te-W-Te layer is weakly bonded to adjacent layers via van der Waals interactions along the $c$-axis (Fig.\,\ref{TopoTeo}(a)), contrasting with the strong covalent bonds within each layer. As depicted in the atomic structures in Fig.\,\ref{FigTopo}(a) and Fig.\,\ref{TopoTeo}(a), the tungsten (W) atoms within a layer form zigzag chains extending along the a-axis. Consequently, scanning tunneling microscopy (STM) topography images reveal tellurium (Te) atoms arranged in chains along this direction. The inset of Fig.\,\ref{FigTopo}(a) shows a high-resolution STM image with the atomic structure overlaid, providing direct correlation between the topography and the atomic lattice. The main panel of Fig.\,\ref{FigTopo}(a) further allows for the identification of Te rows and the presence of point defects such as interstitials and vacancies. These defects, a tungsten interstitial (black arrow) and a tellurium vacancy (white arrow), are more clearly visualized in the inset of Fig.\,\ref{TopoTeo}(b).

To elucidate the topographic features observed around defects, we conducted first-principles calculations using DFT as implemented in the Quantum Espresso code. WTe$_2$ crystallizes in an orthorhombic structure (polytype 1T$_d$) with space group Pnm21 and a primitive unit cell containing 12 atoms. The calculated lattice parameters are $a = 3.477$ \AA, $b = 6.249$ \AA\,and $c = 14.0179$ \AA.  

To correlate the topographic features around defects with the electronic structure, we calculated the local density of states (LDOS) at an energy of 100 meV and compared these results with STM topography acquired at a bias voltage of 100 mV (Figs.,\ref{TopoTeo}(c-f)). High-resolution STM images of a tellurium vacancy and a tungsten interstitial are presented in Figs.,\ref{TopoTeo}(c) and (d), respectively, while the corresponding calculated LDOS maps are shown in Figs.\,\ref{TopoTeo}(e) and (f). Notably, our calculations accurately reproduce the enhanced LDOS along the tellurium chains, which gives rise to the characteristic row-like appearance in WTe$_2$ STM images. This row-like structure is locally disrupted at defect sites. At a Te vacancy (Fig.\,\ref{TopoTeo}(c)), both experiment and calculation (Fig.\,\ref{TopoTeo}(e)) reveal a distinct depression or "hole" in the Te row corresponding to the missing atom. Experimentally, the Te chains appear compressed near this vacancy, a feature that may be related to the perpendicular features observed in the calculated LDOS around the vacancy (Fig.\,\ref{TopoTeo}(e)). Conversely, the calculated LDOS for a tungsten interstitial (substitution of Te with W) in Fig.\,\ref{TopoTeo}(f) shows a significant increase in LDOS with a highly asymmetric, horseshoe-like structure around the substituted W atom. This characteristic feature is also clearly observed in the experimental STM image of a W interstitial (Fig.\,\ref{TopoTeo}(d)). The overall agreement between experimental topography and calculated LDOS for both types of defects indicates that both tungsten and tellurium derived orbitals contribute uniquely to the tunneling current at the atomic scale. This sensitivity to the atomic species is crucial for understanding the spatial variations in the peak sequence observed in the tunneling conductance at high magnetic fields when measuring with atomic resolution.

We now focus on two types of point defects: (i) a tellurium vacancy in the surface layer and (ii) a tungsten antisite defect where a tellurium atom is replaced by a tungsten atom. To model the experimental system, we constructed 4$\times$4$\times$2 supercells containing four layers, with a vacuum separation of 13\,\AA\,between periodic replicas of the defects. This separation is sufficient to minimize spurious interactions between neighboring defect images. We performed first-principles calculations based on spin-polarized DFT within the plane-wave formalism, as implemented in the QuantumESPRESSO package\,\cite{Giannozzi_2009}. The exchange-correlation energy was treated using the generalized gradient approximation (GGA) with the Perdew–Burke–Ernzerhof (PBE) functional\,\cite{PhysRevLett.77.3865}, and core electrons were represented by standard Ultra-soft pseudopotentials (USPP) obtained from the Materials Cloud Database. The electronic wave functions were expanded using plane waves with well-converged kinetic energy cutoffs of 50 Ry for the wave functions and 400 Ry for the charge density. To account for van der Waals interactions between the WTe$_2$ layers, we incorporated semiempirical Grimme-D3 corrections. Atomic positions were optimized using the Broyden-Fletcher-Goldfarb-Shanno (BFGS) algorithm until the residual forces on each atom were less than $10^{-5}$ Ry/au and the energy difference between successive relaxation steps was below $10^{-7}$ Ry. The Brillouin zone was sampled with a dense $\Gamma-$centered 3 $\times$ 2 $\times$ 1 k-point Monkhorst-Pack\,\cite{PhysRevB.13.5188}. To prevent artificial interactions between periodic images along the non-periodic z-direction, a vacuum spacing 18\,\AA\,was introduced. 

\begin{figure*}
	\includegraphics[width=1\textwidth]{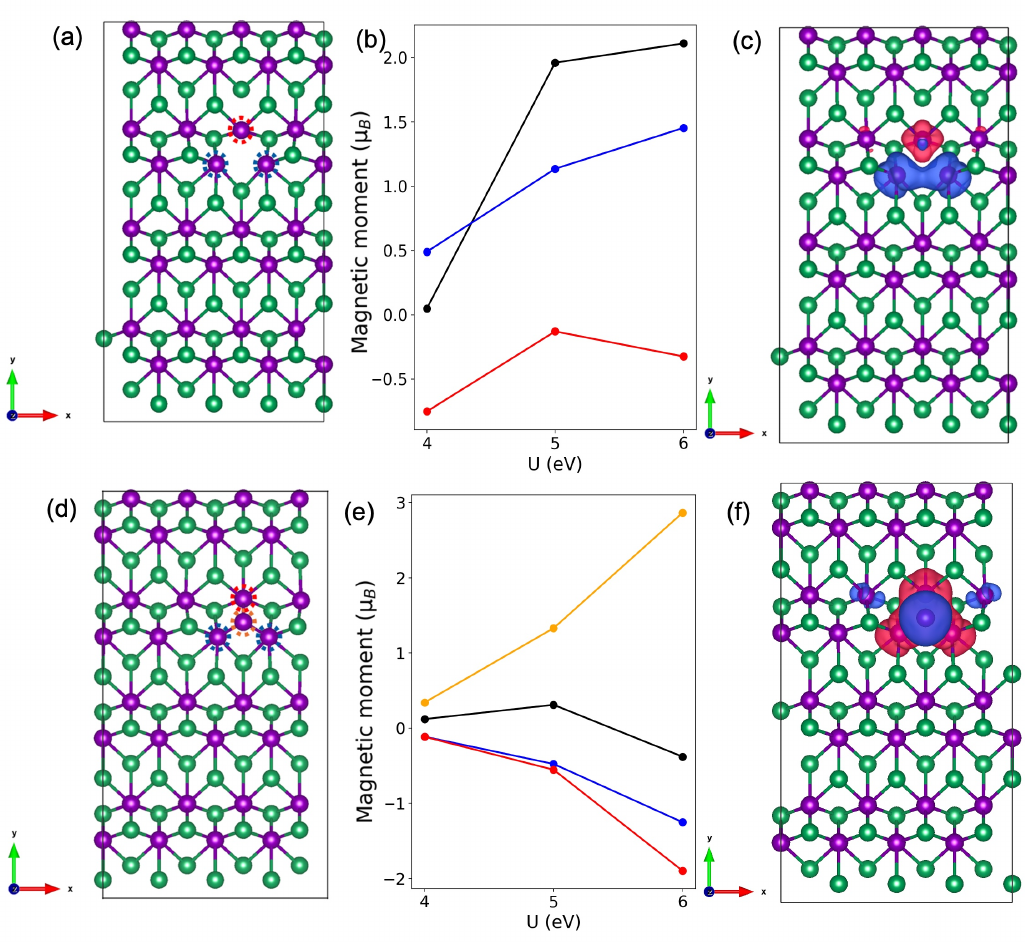}
	\caption{(a) Local atomic structure around a Te vacancy. This panel illustrates the surface atomic arrangement surrounding a tellurium (Te) vacancy. Dashed circles highlight the two distinct types of tungsten (W) atoms in the vicinity of the defect, differentiated by their bonding configurations. (b) Magnetic moment dependence on Hubbard U at the Te vacancy. This plot shows the calculated magnetic moment as a function of the Hubbard U parameter at the specific atomic sites indicated by colored markers in panel (a). Black points represent the average magnetic moment across these sites. The results indicate that the magnetic moments are largely compensated, suggesting antiferromagnetic coupling, except at larger values of U. (c) Spin isosurface around a Te vacancy (U=5 eV). This panel displays a spin isosurface for a Te vacancy, calculated with a Hubbard U parameter of 5 eV. Red regions indicate a spin imbalance of 0.003 (spin down), while blue regions represent a spin imbalance of 0.003 (spin up). (d) Surface atomic arrangement around a W interstitial. Analogous to panel (a), this panel depicts the surface atomic arrangement around a tungsten (W) interstitial defect. (e) Magnetic moment dependence on Hubbard U at the W interstitial. Similar to panel (b), this plot shows the calculated magnetic moment as a function of the Hubbard U parameter at relevant atomic sites near the W interstitial. Black points represent the average moment. (f) Spin isosurface around a W interstitial (U$=$5 eV). Corresponding to panel (c), this panel presents the spin isosurface for a W interstitial, calculated with a Hubbard U parameter of 5 eV. Red and blue regions indicate spin down and spin up imbalances of 0.003, respectively. Tungsten (W) and tellurium (Te) atoms are represented by violet and green circles.}
	\label{TeoMagImp}
\end{figure*}

For the tellurium vacancy defect (Fig.\,\ref{TeoMagImp}(a)), we analyzed two distinct tungsten (W) sites based on their coordination environment. Our calculations revealed an antiferromagnetic correlation between these two W sites across the entire range of the Hubbard U parameter, resulting in a net positive magnetic moment for the defect region (Fig.\,\ref{TeoMagImp}(b)). The spin isosurface representation (Fig.\,\ref{TeoMagImp}(c)) shows a tendency towards a two-fold symmetric arrangement of spin density, a feature that is consistent with the experimental STM topography observed for Te vacancies (Fig.\,\ref{TeoMagImp}(e)).

In the case of the tungsten interstitial defect, there are three inequivalent W sites, distinguished by their coordination (Fig.\,\ref{TeoMagImp}(d)). Our findings indicate that the W atoms surrounding the interstitial exhibit ferromagnetic interactions among themselves, while coupling antiferromagnetically to the interstitial W atom across the studied U range (Fig.\,\ref{TeoMagImp}(e)).

The magnetic density of states we find around defects suggest that scattering leading to QPI patterns could require a careful consideration of eventual magnetic effects in scattering. If, for example, a surface band has magnetic texture, scattering by the impurities considered here might induce changes in the magnetic moment. The influence of these magnetic patterns in QPI requires a complete scattering model which is out of the scope of this work and remains for future studies.

\subsection{Quasiparticle interference}

In the large-scale STM topography image (Fig.\,\ref{FigTopo}(a)), defects are observed across the entire scanned area, yielding an estimated defect density of approximately $1.3 \times 10^{12}$ cm$^{-2}$. Given the lattice parameters $a$ = 3.5 \AA\,and $b$ = 6.25 \AA,  the average distance between defects corresponds to roughly 336 unit cells, resulting in an estimated mean free path $\ell \sim$ 167.5 nm. Applying the Drude model with this mean free path, we calculate a residual resistivity of $\rho = 1.95$\,$\mu\Omega \cdot$ cm and an electron mobility of $\mu=$4.26\,m$^2$/$($Vs$)$. These values are consistent with previously published results and macroscopic characterization of our WTe$_2$ samples\,\cite{wu2015,jo2019}.

The STM topography Fig.\,\ref{FigTopo}(a) also reveals extended features surrounding the point defects. These larger-scale modulations are attributed to the scattering of conduction electrons by the defects, which manifest as the QPI patterns observed in conductance maps acquired at various bias voltages (Fig.\,\ref{FigTopo}(b,c)). These QPI phenomena will be discussed in greater detail in the subsequent section.

\begin{figure*}
	\includegraphics[width=1\textwidth]{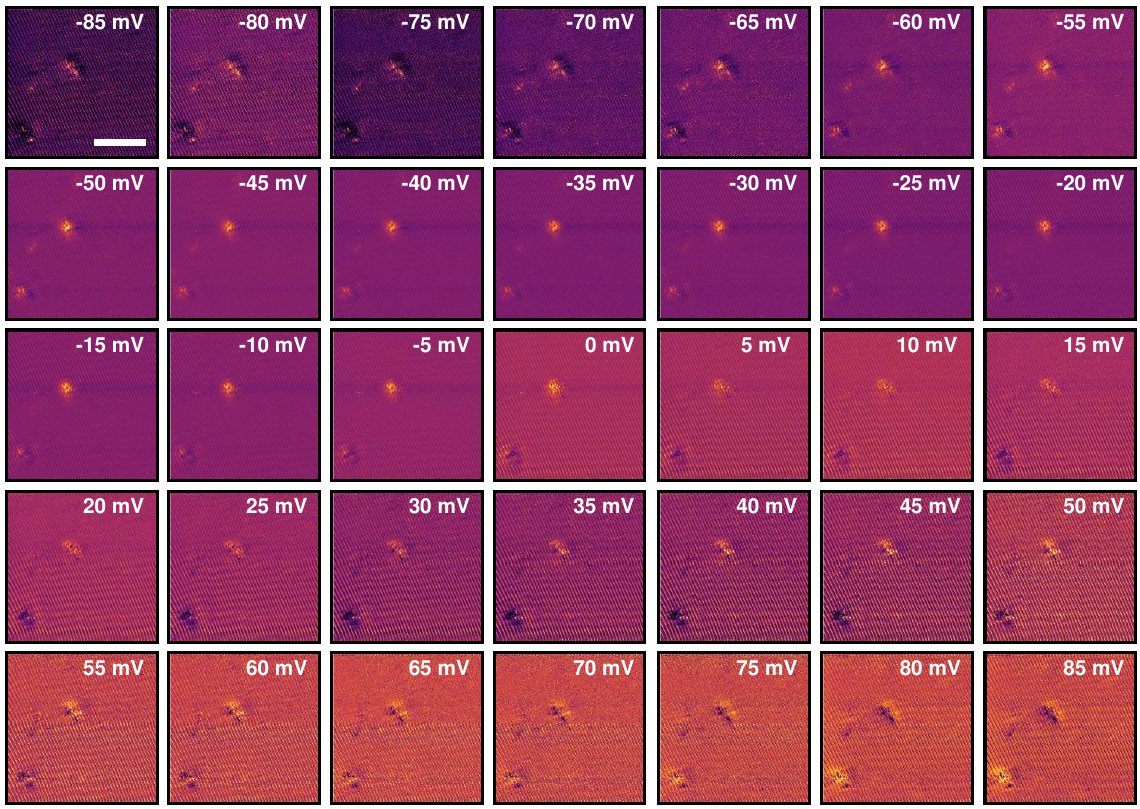}
	\caption{Tunneling conductance maps of the WTe$_2$ surface as a function of the bias voltage. The bias voltage is given in each panel in mV. The white horizontal scale bar in the top left panel represents 10 nm and applies to all maps.}
	\label{ConducMaps}
\end{figure*}

The tunneling conductance is proportional to the local density of states LDOS. The LDOS is related to the electrons wavefunction $\Psi (\vec{r}_k)$ by
\begin{align}
	LDOS (E,\vec{r}) \propto \sum_{\vec{k}} |\Psi(\vec{r}_k)|^2 \delta \left(E-\epsilon (\vec{k})\right),
\end{align}
where $\epsilon(\vec{k})$ is the dispersion relation. When the periodicity of the crystal is broken by the presence of impurities or defects on the surface, electrons are scattered, producing oscillations in the LDOS. These scattering processes are typically elastic. In this picture, scattering between states with $\vec{k}_1$ and $\vec{k}_2$ gives rise to a modulation with $\vec{q} = \vec{k}_1 - \vec{k}_2$ in the LDOS that can be observed in the tunneling conductance with the STM. The scattering probability between an initial state, $i$, and a final state, $f$, is described by the Fermi golden rule,
\begin{align}
\omega (i\longrightarrow f) \propto \frac{2\pi}{\hbar} |V(\vec{q})|^2 N_i(E_i, \vec{k}_i) N_f (E_f, \vec{k}_f),
\end{align}
where $E_i = E_f$ for elastic scattering, $\vec{q} = \vec{k}_f - \vec{k}_i$ is the scattering vector, $N_i$ and $N_f$ the initial and final density of states, and $V(\vec{q})$ the scattering potential. The LDOS$(E,\vec{r})$ is then proportional to the scattering intensity, which is maximal when $\vec{q} = \vec{k}_1 - \vec{k}_2$ joins portions of the band structure with a large density of states, $N_1(E, \vec{k}_1)$ and $N_2(E, \vec{k}_2)$. The Fourier transform of the LDOS$(E,\vec{r})$, LDOS$(E,\vec{q})$, is then related to the autocorrelation function of the band dispersion and we can write LDOS$(E,\vec{q})\propto \sum_{\vec{k}}N(E, \vec{k})N(E, \vec{k}+\vec{q})$. The autocorrelation is weighted by broadening, impurity scattering, spin and orbital effects.

To obtain the scattering intensity from calculation, we first performed DFT calculation in the way described above. After the first-principles calculation, we constructed the Wannier functions~\cite{Marzari1997,Souza2001} using the \textsc{Wannier90} software~\cite{Pizzi2020}.
We did not perform the maximal localization procedure to prevent the mixture of the different spin components.
We took the W-$d$ and Te-$p$ orbitals as the Wannier basis.
Fermi surfaces and autocorrelation functions were calculated using the tight-binding model consisting of the Wannier functions. The autocorrelation function calculated in this study was defined as
\begin{equation}
I(E, \vec{q}) = \frac{1}{2}\sum_{\vec{k}, i, j} \sum_{l=0}^3 f_l(\vec{k}, \epsilon_i - E) f_l(\vec{k}+\vec{q}, \epsilon_j - E)
\label{eq:QPI_theory}
\end{equation}
where $f_l(\vec{k},\epsilon_i - E)$ is the Lorentzian function,
\begin{equation}
f_l(\vec{k}, \epsilon_i - E) = \frac{1}{\pi}\frac{\gamma}{(\epsilon_i - E)^2 + \gamma^2} \langle \sigma_l \rangle_i ,
\end{equation}
with the smearing width $\gamma$ for the $i$-th Kohn-Sham energy $\epsilon_i$ and the expectation value of the Pauli matrix $\sigma_l$ ($l=0,1,2,3$) for the corresponding eigenstate (see, e.g., Ref.~\cite{Inoue2016}, for a use of the spin-decomposed spectral density).
To reduce computational cost, we constrained the summation on the $k_z=0$ plane.
We took $\gamma = 1$ meV and a $240\times 240$ ${\bm k}$-mesh in the region of $|k_{x,y}|\leq 0.6 \pi a^{-1}$.
The Fermi energy in all the calculated data was shifted by $-45$ meV so as to reproduce the experimental data.

\begin{figure*}
	\includegraphics[width=1\textwidth]{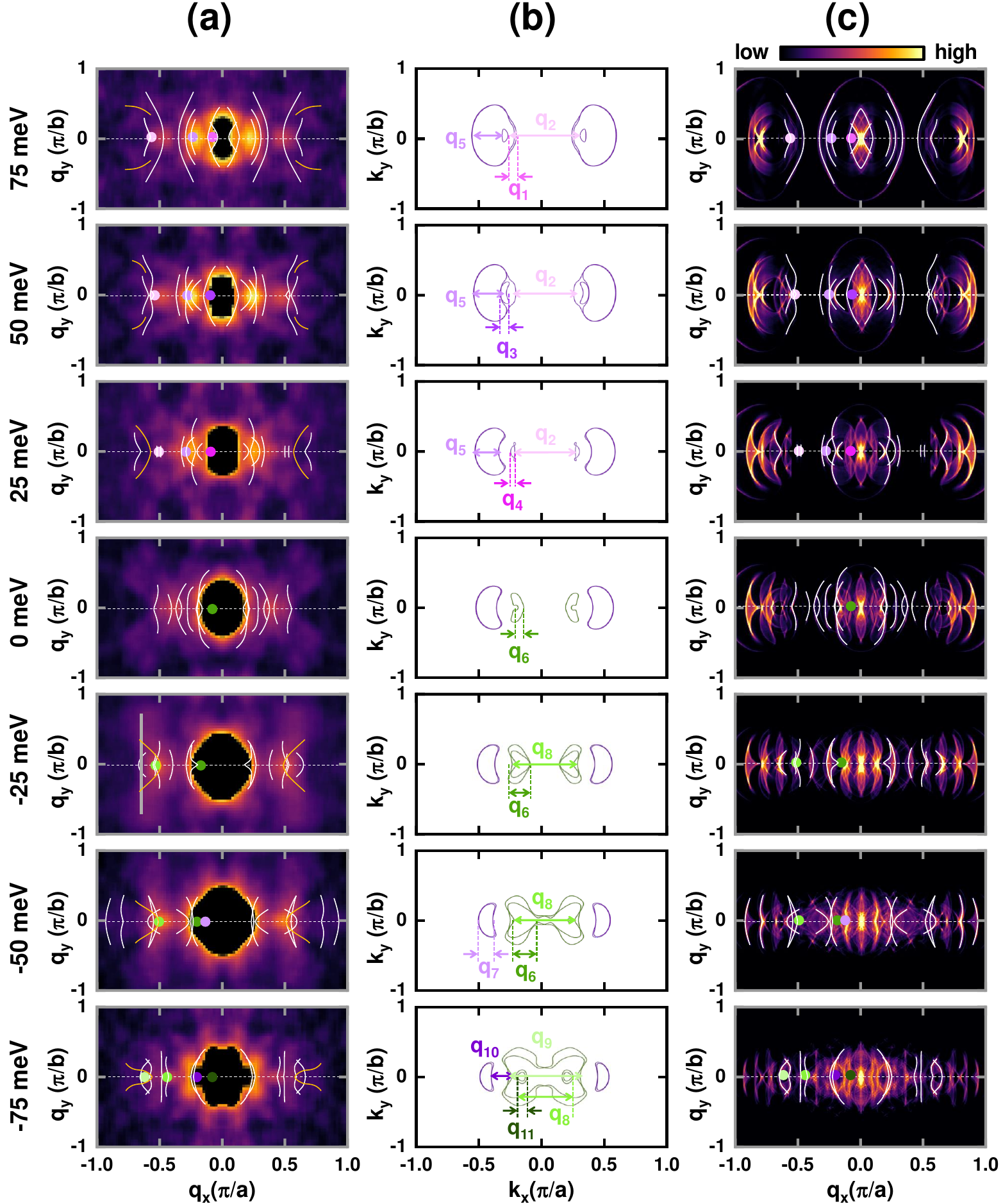}
	\caption{Fourier transforms of conductance maps near defects. These panels display the Fourier transforms of conductance maps acquired over a field of view containing various defects, including both interstitials and vacancies (Fig.\,\ref{ConducMaps}). The corresponding bias voltage (in mV) is indicated to the left of each panel. For clarity, only a selection of bias voltages is shown. Colored points highlight the scattering features in the Fourier transforms that correspond to the scattering vectors $q_1$ to $q_{11}$ identified in the $k_y =0$ line cuts of Fig.\,\ref{FigTopo}(b,c). (b) Calculated bulk band structure constant energy contours and scattering vectors. Calculated constant energy contours of the bulk electronic band structure are shown as colored lines. Colored arrows overlaying these contours represent the scattering vectors associated with the colored points in panel (a), using the same color code for direct correspondence. A schematic representation of the most prominent features of the calculated surface states is also provided in orange. (c) Autocorrelation of calculated constant energy contours. This panel shows the autocorrelation of the calculated bulk band structure constant energy contours. Relevant scattering wavevectors derived from this autocorrelation are highlighted as colored points. White lines in both (a) and (c) represent the bulk band structure constant energy contours at energies close to those of the colored points. The grey line in (a) (third panel from the bottom) is the same $q_y$ cut that is analyzed in Fig.\,\ref{FigQuantumnumber}(e,f).}
		\label{QPIComparison}
\end{figure*}

In Fig.\,\ref{ConducMaps} we show the conductance maps of an area with both vacancy-like and interstitials defects. We used the Fourier transform of these maps to study the QPI scattering, and obtain the scattering profile as a function of the bias voltage shown in Fig.\,\ref{FigTopo}(b) and Fig.\,\ref{QPIComparison}(a) here.

We show in Fig.\,\ref{QPIComparison}(a) the Fourier transform of the tunneling conductance maps, compared to calculations of the bulk band structure (Fig.\,\ref{QPIComparison}(b,c)). In Fig.\,\ref{QPIComparison}(b) we show the constant energy contours of the bulk band structure and in Fig.\,\ref{QPIComparison}(c) their autocorrelation function.

We now turn to a comparison between the bulk band structure and our experimental QPI data. In Figs.\,\ref{QPIComparison}(a) and (c), we plot the $q_y=0$ QPI features (corresponding to scattering along the $a$-direction, as shown in Fig.\,\ref{FigTopo}) as colored points, and the corresponding calculated scattering wavevectors are marked in Fig.\,\ref{QPIComparison}(b). Here, we focus on the QPI behavior away from the $a$ direction where $q_y\neq0$. Starting at a bias voltage of 75 meV (upper panels of Figs.,\ref{QPIComparison}(a) and (c)), the largest wavevector, $q_2$ corresponds to a small S-shaped feature that is consistently observed in both experimental QPI maps and our band structure calculations. Adjacent to this, we find a convex feature that also shows good agreement between experiment and theory. Notably, this convex feature diminishes in intensity as the bias voltage is decreased.

The other two calculated wavevectors, $q_5$ and $q_1$, correspond to concave features that are also present in both our experimental QPI maps and band structure calculations. However, at a bias voltage of $+50$ mV, while the calculations predominantly predict concave features, the experimental QPI clearly reveals several prominent convex features. We attribute this discrepancy to the influence of the surface electronic band structure (plotted for clarity for a single band in Figs.\,\ref{QPIComparison}(a)), which we discuss in more detail below.

When further decreasing the bias, at 0\,meV we see that the scattering intensity is reduced, in experiment and calculations. In experiment we see a few scattering wavevectors of small size. Close to $q_6$ we find an intricate set of features that provide some concave like feature observed in the experiment. For clarity, we do not show these features in the panels below zero bias in Fig.\,\ref{QPIComparison}(a). Instead, for the panels with negative bias, we have masked the features close to zero wavevector by enlarging the corresponding filter. Accordingly, the central dark area in the bottom panels of Fig.\,\ref{QPIComparison}(a) is larger. Nevertheless, we mention that, for small $q_y$ values, the experimental QPI features closely match the calculated scattering wavevectors found for $q_y=0$ across the negative bias range. In the vicinity of $q_6$, a concave feature is prominent at -25 meV, which evolves into an X-shaped feature at -50 meV, again showing good agreement between experiment and calculation. Furthermore, near $q_8$ we observe convex and S-like features in both the experimental data and the band structure calculations.

By increased masking of features for low wave vectors, we can now see better the convex features mentioned above. We associate these with the surface band structure as these are absent in the bulk band structure scattering patterns. For clarity, we only plot in orange a relevant surface band in Figs.\,\ref{QPIComparison}(a) for negative bias voltages. The grey line in the -25 meV panel indicates the cut along $q_y$ that was analyzed in Fig.\,\ref{FigQuantumnumber}(e,g). We note that there is a certain coincidence between the observed scattering pattern and the expected pattern from surface bands. The large convex feature observed experimentally between zero bias and \mbox{-75 mV} is absent in the bulk electronic band structure, but is close to the orange convex shape in the three bottom panels of Fig.\,\ref{QPIComparison}(a)), albeit strong mixing with other scattering patterns.

ARPES studies typically focus on higher binding energies and occupied electronic states, whereas our STM technique allows us to probe the band structure with finer energy resolution, including unoccupied states above the Fermi level\,\cite{pletikosic2014}. Overall, there is good agreement between ARPES and STM findings obtained here and in Refs.\,\cite{zhang2017,yuan2018,Hwang2020}. ARPES is particularly sensitive to the hole-like pockets originating from tellurium orbitals and their structure well below the Fermi level. High-resolution ARPES measurements have also revealed the same convex features in these hole pockets at small $k_x$ that we observe in our QPI data\,\cite{wu2015,wu2016}. Notably, ARPES has identified a linearly dispersing surface band that disperses along $k_y$ at a similar energy and momentum range to the feature we discuss in Fig.\,\ref{FigQuantumnumber}\,\cite{wu2015,wu2016}. This linear dispersion roughly follows the shape discussed in Fig.\,\ref{FigQuantumnumber}(g,h).

We note that there is a weak feature which is not marked in the Fig.\,\ref{FigTopo}(b) for bias voltages between about 50 mV and 80 mV and large $q_x$. We did not mark it on purpose in the main text, for clarity. The feature shows a dispersion with energy such that it shifts to smaller $q_x$ for higher bias. The same feature is seen as a weak signal with a concave shape in the top panel of Fig.\,\ref{QPIComparison}(a). There is indeed another scattering wave vector which provides very intense patterns in the calculations for the three top panels of Fig.\,\ref{QPIComparison}(c) for large $q_x$. This corresponds to scattering between the small inner pockets at negative $k_x$ and the outer sides of the large pockets at positive $k_x$.

\subsection{Parameters used in the calculation of the Landau level sequence}

To generate the data presented in Fig.\,\ref{FigQuantumnumber}(d) we consider a parabolic hole band dispersing along $k_y$ with its band top positioned at -13 meV. For the linear dispersing band, we utilize an electron band and place the band bottom at -30 meV. These two choices are inspired by the calculations of the surface band structure, Fig.\,\ref{FigQuantumnumber}(g), but are of course a considerable simplification. We do not know of an expression providing the Landau level sequence of a semi-linear band. We thus assume that, instead of the usual Dirac sequence, given by $N(E>E_{0,lin})=\sum_m \delta(\vert E-E_{0,lin}\vert +\sqrt{2m\hbar e B v_F^2})$ we can use $N(E>E_{0,lin})=\sum_m \delta(\vert E-E_{0,lin}\vert +\sqrt{2\hbar e B v_F^2(m+0.3m^2)})$. This modification only acts far from the bottom of the band and leads to equally spaced Landau levels when the linear dispersing surface band merges with other bands (positive energies in Fig.\,\ref{FigQuantumnumber}(d)). Additionally, we incorporate Lorentzian broadening to the Landau levels and Gaussian broadening to the tunneling conductance to simulate experimental conditions. The resulting density of states as a function of energy, displaying the superposition of Landau level peaks from both bands, is shown in Fig.\,\ref{FigQuantumnumber}(d). Notably, the Landau level peaks exhibit distortion near the crossing point due to the proximity and subsequent merging (due to broadening) of levels originating from the parabolic and linear dispersing bands. With these parameters, the calculated Landau level sequence shows good agreement with the experimental results presented in Fig.\,\ref{FigQuantumnumber}(b).

\subsection{Spatially averaged tunneling conductance vs magnetic field}

\begin{figure*}
	\includegraphics[width=1\textwidth]{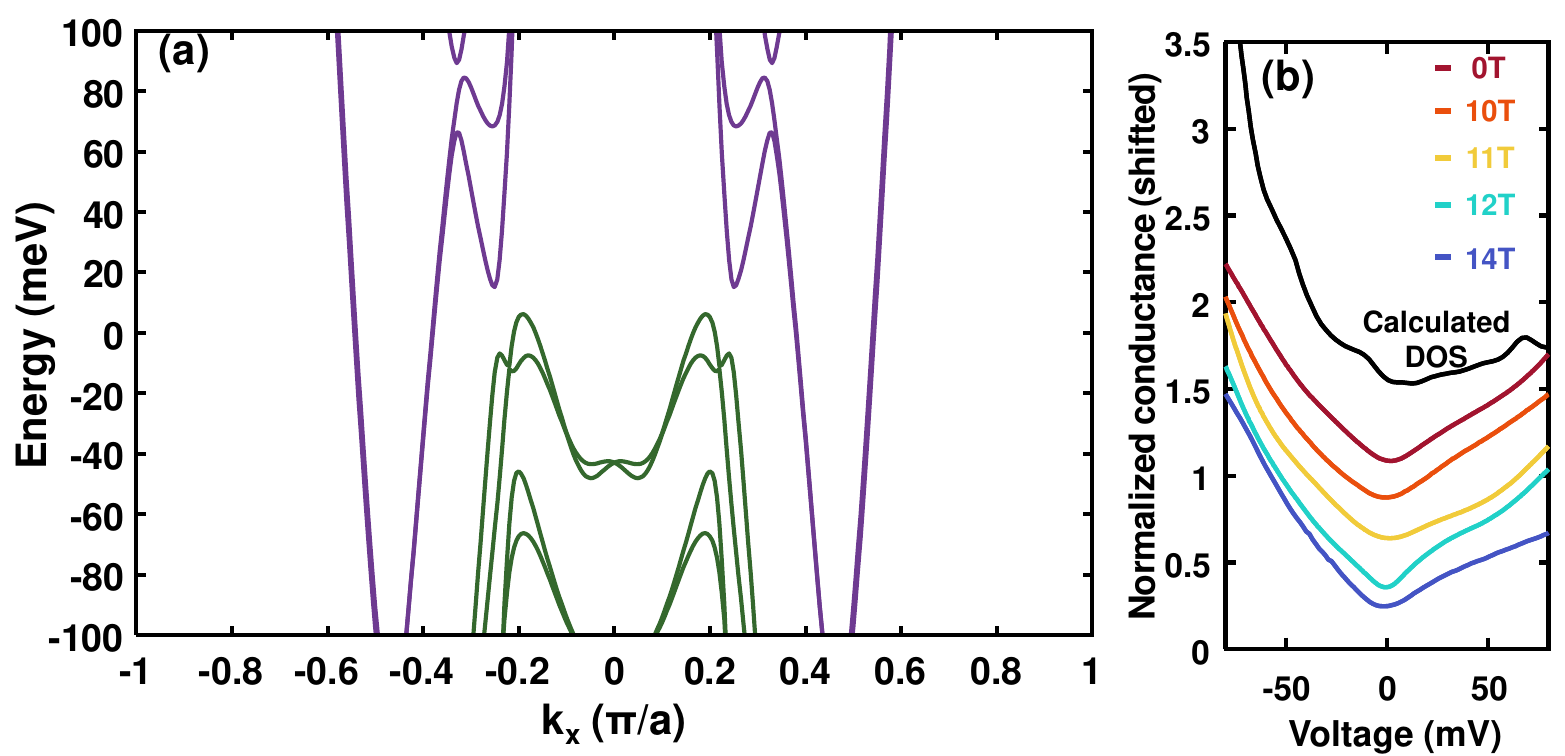}
	\caption{(a) Calculated bulk band structure along the $\Gamma$ ($k_x$ = 0) to $X$ ($k_x$ = 1) direction inside the energy range explored in the experiment. Electron and hole bands are depicted in purple and green, respectively. Electron bands are derived from W 5d-orbitals and hole bands from Se 5p-orbitals. The calculations are shifted 45\,meV downwards in energy to match our experimental results. (b) Spatially averaged normalized tunneling conductance versus bias voltage curves measured at magnetic fields ranging from 0\,T to 14\,T (colored lines). The normalized density of states obtained from the band structure in (a) is also plotted as a black line. Curves are shifted vertically for better visualization.}
		\label{BandStructure}
\end{figure*}

Fig.\,\ref{BandStructure}(a) presents our DFT band structure calculations along the $\Gamma - X$ direction, covering the same energy range as our STM measurements. Electron bands, with a dominant tungsten 5d orbital character, are depicted in purple, while hole bands, primarily of tellurium 5p orbital character, are shown in green. The resulting momentum-integrated density of states derived from these bands is plotted as a black line in Fig.\,\ref{BandStructure}(b). We compare this calculated density of states to the spatially averaged tunneling conductance measured at various magnetic field strengths (colored lines in Fig.\,\ref{BandStructure}(b)). Notably, the overall shape of the experimental spatially averaged tunneling conductance curves as a function of bias voltage closely resembles the energy dependence of the calculated density of states.

STM probes the local electronic structure through the quantum mechanical tunneling of electrons between the sample and the atomically sharp tip, a process governed by the overlap of their respective wavefunctions\,\cite{PhysRevB.31.805}. This tunneling probability is not uniform across different atomic sites on the sample surface, as it involves varying contributions from different portions of the material's band structure\,\cite{PhysRevB.95.100502,PhysRevB.77.134505,PhysRevLett.92.206101,doi:10.1126/sciadv.aao0362}. As illustrated in our topographic image (Fig.\,\ref{TopoTeo}(a)), we clearly resolve tellurium (Te) atomic rows with distinct contrast observed between adjacent Te sites. This suggests that tunneling primarily occurs through tellurium-derived electronic bands when the STM tip is positioned directly above Te atoms, and predominantly through tungsten-derived bands when the tip is located in the interstitial regions between Te sites. Consequently, when tunneling into electronically distinct atomic sites, the specific set of bands contributing to the LDOS is altered, leading to variations in the energy positions of maxima and minima observed in the tunneling conductance spectra.

A key question arises: how can the resistance in WTe$_2$ increase by over six orders of magnitude upon application of a magnetic field, without significant magnetic field-induced alterations to its electronic band structure? Our data show that the remarkable colossal magnetoresistance observed in this material is not driven by changes in the Fermi surface nor the band structure away from the Fermi energy induced by the magnetic field. Instead, our data strongly supports the proposal of Refs.\,\cite{ali2014,PhysRevLett.114.156601}, which attributes the large magnetoresistance to a near-perfect compensation between the densities of electrons and holes.

\begin{figure}
	\includegraphics[width=1\columnwidth]{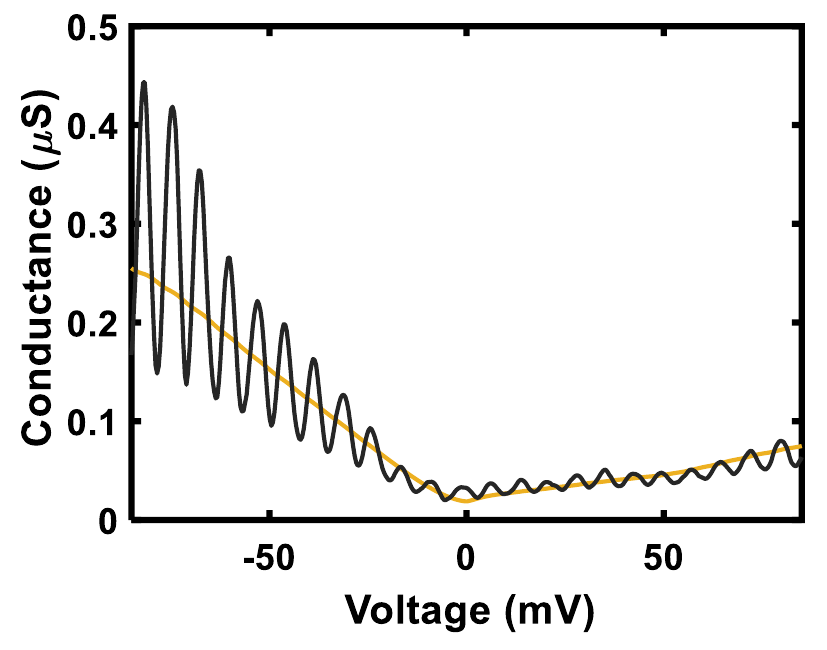}
	\caption{We show as a brown line the spatially averaged tunneling conductance. Black line shows the tunneling conductance obtained at an atomically well defined position. The dependence as a function of position is discussed in Fig.\,\ref{FigConductanceMap}. Data are taken at a magnetic field of 14 T.}
	\label{Background}
\end{figure}

The observation of Landau quantization occurs  at atomic scale and is discussed in the main text and in the following. To obtain the data of Fig.\,\ref{FigLandau} we have used the spatial average of the field of the tunneling conductance in the field of view where we acquired those data. In Fig.\,\ref{Background} we show with a black line the obtained tunneling conductance at atomic scale and with a brown line the background used for subtraction. We add an additional bias dependent offset to obtain values above zero in Fig.\,\ref{FigLandau}.

We note that we observe a similar behavior for $n$ as the one shown in Fig.\,\ref{FigQuantumnumber}(b) for a magnetic field of 12 T, with a shift in $n'$ in the same bias voltage range. At lower magnetic fields, the amplitude of the peaks in the tunneling conductance is too small to follow the bias dependence of $n$.

\subsection{Atomic scale maps of Landau quantized tunneling conductance}

On the other hand, another relevant question is how do macroscopic experiments as resistivity show quantum oscillations, whereas the spatially averaged tunneling conductance shows no signature of Landau quantization?

\begin{figure*}
	\includegraphics[width=1\textwidth]{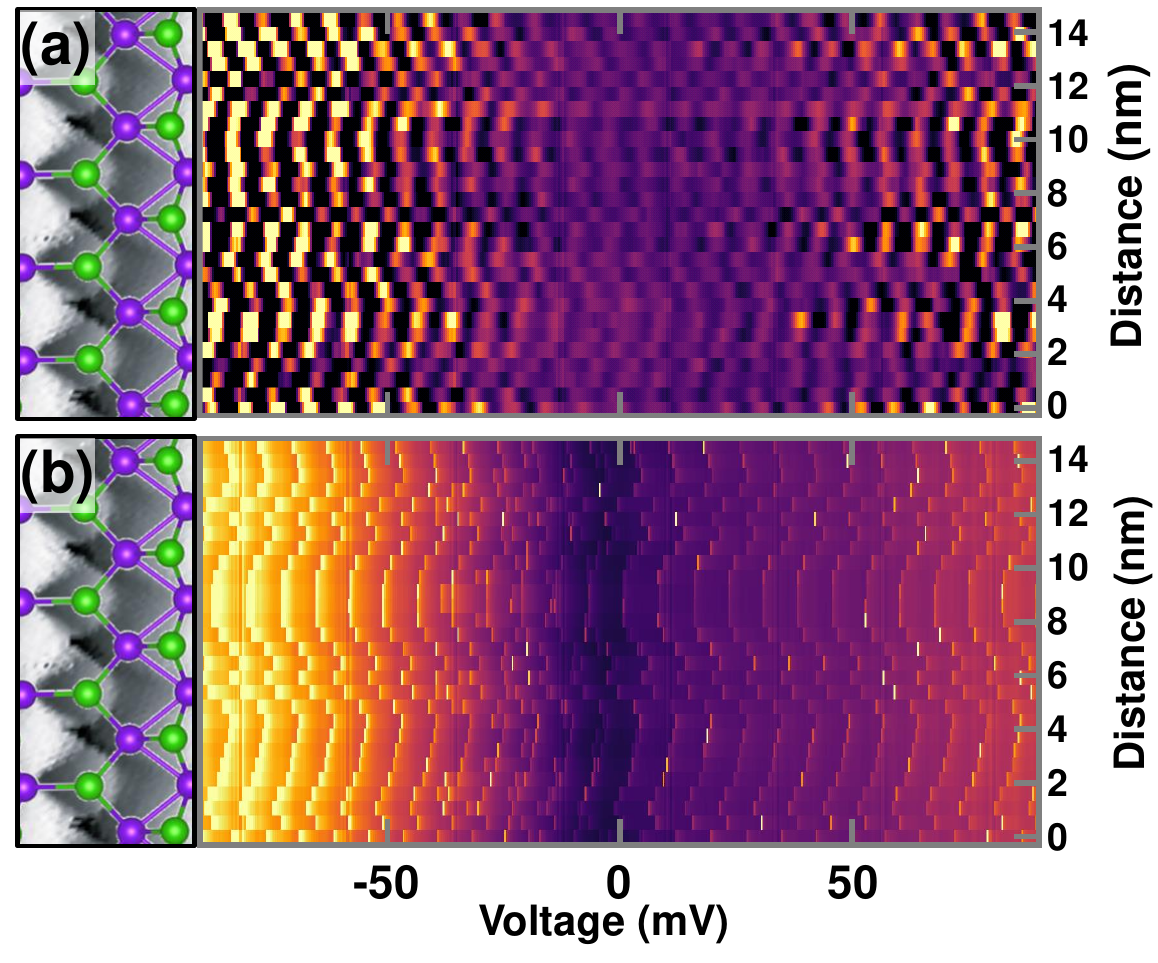}
	\caption{(a) Tunneling conductance along a row following the a axis, taken at 14\,T. In the left panel we show a topography of the surface, scaled to match the distance and atomic locations in the y-scale of the main panel. The map was obtained on top of a Te row. (b) Same as in (a), with a map obtained calculating the density of states as explained in the text, with parameters $A = 4.2$\,mV and $q =2\pi/a$. We use the same pixel size as in the experiment and the level separation measured for 14\,T.}
	\label{FigConductanceMap}
\end{figure*}

Let us first address the effect of atomic size changes on the spatial average of the tunneling conductance. Landau quantization is observed in WTe$_2$ with STM when taking tunneling conductance curves with atomic precision. We show in Fig.\,\ref{FigConductanceMap} the tunneling conductance as a function of the position along a line on top of Te atoms. We observe that the oscillations due to Landau levels shown in Fig.\,\ref{FigLandau} change as a function of the position (Fig.\,\ref{FigConductanceMap}(a)). We can obtain a similar map taking a spatially varying density of states $N_{local} (E,x) = \sum_m \left[E-E_0 - \sqrt{2\hbar e B v_F^2(m+0.3m^2)} + A\cos(qx)\right]^{-\frac{1}{2}}$. We use  $A$ = 4.2 mV, $v_F=2.8\times 10^5$ $m/s$ and $q = 2\pi a$, where $a$ is the unit cell parameter in the direction of the Te chains. Note that $A$ is larger than $\hbar\omega_c/2$. Hence, the result is not a simple cosine wave, but a spatially modulated set of maxima and minima, as shown in Fig.\,\ref{FigConductanceMap}(b). The oscillations are very rapid, and are averaged to zero after a few lattice constants.

Within this minimal model, the atomic scale shift in the peak sequence is associated to band selective tunneling and a shift in the Landau level sequence on different tunneling positions.

\subsection{Comparison with macroscopic experiments}

Even within the considerations given in the previous section, a spatially averaged tunneling conductance showing no sign of Landau quantization seems difficult to reconcile with the strong quantum oscillations observed in macroscopic experiments. To better explain this, we need to carefully consider the effects of Landau quantization on the two-dimensional density of states at the surface and on the whole three-dimensional bandstructure.

The quantum oscillations observed in resistivity and in macroscopic probes of the Fermi level density of states are due to the passage of Landau tubes through extremal areas of the Fermi surface. As such, any experiment measuring the density of states at the Fermi level as a function of the magnetic field is expected to present an oscillatory behavior. However, we are not measuring the density of states at the Fermi level as a function of the magnetic field  $N(E=E_F,B)$. Instead, we are measuring the density of states as a function of energy at a fixed magnetic field $B_0$, $N(E,B=B_0)$. The latter leads to the peaks observed in the tunneling conductance as a function of the bias voltage observed in many previous experiments\,\cite{doi:10.1126/science.1171810,doi:10.1126/science.aag1715,PhysRevB.82.081305,doi:10.1126/science.1239451,PhysRevB.107.115426,PhysRevLett.118.016803,Tsui2024,Yin2016,PhysRevB.92.165420}.

STM is of course sensitive to $N(E=E_F,B)$. However, by the very nature of the measurement, the current is fixed by the feedback loop when changing the magnetic field $B$. This implies that the tip height adjusts to the variations in the tunneling conductance induced by the magnetic field changes in $N(E=E_F,B)$. Therefore, we can expect that the tip height presents an oscillatory behavior as a function of the magnetic field which should reproduce the bulk oscillatory behavior in macroscopic experiments. It is, however, a matter of fact, that this behavior is much more tricky to observe in usual STM experiments than measuring at a fixed magnetic field the tunneling conductance as a function of the bias voltage, which is proportional to $N(E,B=B_0)$. The influence of the magnetostriction or noise induced by Foucault currents makes it very difficult to resolve the quantum oscillations in magnetic field sweeps. Furthermore, STM is a local measurement, which requires a large area free of impurities during the whole field sweep, something which might be difficult to obtain due to the magnetostriction of tip, sample and the whole set-up. Of course, it is not impossible and it might be achieved in future work, but it is not within usual STM capabilities and the technique has not been developed.

On the other hand, we should also note that we resolve particularly the surface when measuring $N(E,B=B_0)$. As explained in the main text, Landau quantization as a function of the energy is easier to resolve for a two-dimensional band structure. This provides a sensitivity to surface states which is absent in macroscopic experiments. If a technique to succesfully measure $N(E=E_F,B)$ by STM is developed, we anticipate that it would probably address mostly bulk features of the band structure.

%

\end{document}